\documentstyle[12pt,epsf]{article}
\newcommand {\beq}{\begin{equation}}
\newcommand {\eeq}{\end{equation}}
\newcommand {\beqa}{\begin{eqnarray}}
\newcommand {\eeqa}{\end{eqnarray}}
\newcommand {\n}{\nonumber \\}

\begin{document}
\setlength{\oddsidemargin}{0cm}
\setlength{\baselineskip}{7mm}

\begin{titlepage}
 \renewcommand{\thefootnote}{\fnsymbol{footnote}}
$\mbox{ }$
\begin{flushright}
\begin{tabular}{l}
KEK-TH-669 \\
KUNS-1628 \\
Jan 2000
\end{tabular}
\end{flushright}

~~\\
~~\\
~~\\

\vspace*{0cm}
    \begin{Large}
       \vspace{2cm}
       \begin{center}
         {Bi-local Fields in Noncommutative Field Theory}      \\
       \end{center}
    \end{Large}

  \vspace{1cm}

\begin{center}
           Satoshi I{\sc so}$^{1)}$\footnote
           {
e-mail address : satoshi.iso@kek.jp},
          Hikaru K{\sc awai}$^{2)}$\footnote
           {
e-mail address : hkawai@gauge.scphys.kyoto-u.ac.jp}{\sc and}
           Yoshihisa K{\sc itazawa}$^{1)}$\footnote
           {
e-mail address : kitazawa@post.kek.jp}

        $^{1)}$ {\it National Laboratory for High Energy Physics (KEK),}\\
               {\it Tsukuba, Ibaraki 305, Japan} \\
        $^{2)}$ {\it Department of Physics, Kyoto University,
Kyoto 606-8502, Japan}\\
\end{center}

\vfill

\begin{abstract}
\noindent
\end{abstract}
We propose a bi-local representation in noncommutative
field theory.
It provides a simple description for high momentum
degrees of freedom. It also shows that
the low momentum modes can be well approximated
by ordinary local fields. Long range  interactions
are generated in the effective action for the lower momentum
modes
after integrating out the high momentum bi-local fields.
The low momentum modes can be represented by diagonal blocks
in the matrix model picture
and the high momentum bi-local fields correspond to off-diagonal
blocks.
This block-block interaction picture simply reproduces the infrared singular
behaviors of
nonplanar diagrams in noncommutative field theory.
\vfill
\end{titlepage}
\vfil\eject

\section{Introduction}
\setcounter{equation}{0}
Superstring theory is a leading candidate for the unification of
fundamental interactions. As such it is expected to reconcile
quantum
mechanics and general relativity.
As is well known, field theoretic approach to quantize gravity
encounters serious difficulties at short distances.
String theory contains the notion of the minimal length
(string scale) which is expected to cure such difficulties.

Although the first quantization of string theory is well understood,
it is ripe to obtain fully nonperturbative formulation
of superstring theory. We have proposed such a formalism
as a large $N$ reduced model with the maximal chiral SUSY
\cite{IKKT}. It is a finite theory which has a potential
to predict the structure of space-time\cite{AIKKT}\cite{review}.

Another school of thought which also introduces the minimum
length scale advocates to replace Riemannian geometry by
noncommutative geometry\cite{Connes}.
Furthermore string theory in noncommutative tori
has been studied in \cite{CDS}.

Remarkably, these various schools of thought have converged
recently. We have pointed out that noncommutative Yang-Mills
theory is naturally obtained in IIB
matrix model by expanding the
theory around a noncommutative space-time\cite{AIIKKT}.
We have further studied the Wilson loops. These
investigations have shown
that the theory contains not only point like field theoretic
degrees of freedom but also open string like extended objects
\cite{IIKK}.
In terms of the matrix representation, the former is close
to diagonal and the latter is far off-diagonal degrees of freedom.
In string theory these backgrounds are interpreted as $D$-branes
with constant $b_{\mu\nu}$ field background\cite{SW}.
These problems are further studied in
\cite{AMNS}\cite{bars99}.
\par
In this paper, we continue our investigation of noncommutative
field theory as twisted reduced models.
We have mapped the twisted reduced model onto the
noncommutative field theory by expanding the matrices by
the momentum eigenstates.
In this paper we propose  a more natural decomposition rule for the matrices
in terms of a bi-local basis.
We show that they corresponds to
`open strings'
and we can reproduce quenched reduced models
with long `open strings'.
We also show that the diagonal elements
represent ordinary plane waves.
Noncommutative field theory exhibits crossover at the
noncommutative scale. At large momentum scale,
it is identical to large $N$ field theory
and the relevant degrees of freedom are described by
`open strings' longer than the noncommutative scale
while
it reduces to field theory in the opposite limit.
We make these statements more precise in this paper.
The equivalence to large $N$ field theory at large
momentum scale is further supported
since we find quenched reduced models in noncommutative field theory.
In the low energy limit, we find long range interactions
in noncommutative field theory which are absent in the
ordinary field theory.
We point out that this effect can be simply understood
in terms of block-block interactions in the matrix model
picture.

\section{Noncommutative field theories as twisted reduced models}
\setcounter{equation}{0}

In this section, we briefly recapitulate our approach
to noncommutative field theory as twisted reduced models.
Reduced models are defined by the dimensional reduction of $d$
dimensional gauge theory down to zero dimension (a point)\cite{RM}.
The application of reduced models to string theory is pioneered in
\cite{Zacos}\cite{Bars}.
We consider $d$ dimensional $U(n)$ gauge theory coupled to
adjoint matter as an example:
\beq
S=-\int d^dx {1\over g^2}Tr({1\over 4}[D_{\mu},D_{\nu}][D_{\mu},D_{\nu}]
+{1\over 2}\bar{\psi}\Gamma _{\mu}[D_{\mu},\psi ]) ,
\eeq
where $\psi$ is a Majorana spinor field.
The corresponding reduced model is
\beq
S=- {1\over g^2}Tr({1\over4}[A_{\mu},A_{\nu}][A_{\mu},A_{\nu}]
+{1\over 2}\bar{\psi}\Gamma _{\mu}[A_{\mu},\psi ]) .
\eeq
Now $A_\mu$ and $\psi$ are $n\times n$ Hermitian matrices
and each component of $\psi$ is $d$-dimensional Majorana-spinor.
IIB matrix model is obtained when $d=10$ and $\psi$ is
further assumed to be Weyl-spinor as well.
Any noncommutative field theory is  realized as well
in terms of a twisted reduced model by the same mapping rules
as below but here we explain the case of the above reduced models
of gauge theory.
We expand the theory around the following classical solution:
\beq
[\hat{p}_{\mu},\hat{p}_{\nu}]=iB_{\mu\nu} ,
\eeq
where  $B_{\mu\nu}$ are $c$-numbers.
We assume the rank of $B_{\mu\nu}$
to be $\tilde{d}$ and define its inverse $C^{\mu\nu}$ in $\tilde{d}$
dimensional subspace.
The directions orthogonal
 to the subspace
is called the transverse directions.
$\hat{p}_{\mu}$ satisfy the canonical commutation relations and
they span the $\tilde{d}$ dimensional phase space.
The semiclassical correspondence shows that the
volume of the phase space is $V_p=n(2\pi)^{\tilde{d}/2} \sqrt{detB}$.

We expand $A_{\mu}=\hat{p}_{\mu}+\hat{a}_{\mu}$. We Fourier decompose
$\hat{a}_{\mu}$ and $\hat{\psi}$ fields as
\beqa
\hat{a}&=&\sum_k \tilde{a}(k) exp(iC^{\mu\nu}k_{\mu}\hat{p}_{\nu}) ,\n
\hat{\psi}&=&\sum_k \tilde{\psi}(k) exp(iC^{\mu\nu}k_{\mu}\hat{p}_{\nu}) ,
\label{twist}
\eeqa
where $exp(iC^{\mu\nu}k_{\mu}\hat{p}_{\nu})$ is the eigenstate of
adjoint $P_{\mu}=[\hat{p}_{\mu},~]$ with the eigenvalue $k_{\mu}$.
The Hermiticity requires that $\tilde{a}^* (k)=\tilde{a}(-k)$ and
$\tilde{\psi} ^*(k)=\tilde{\psi} (-k)$.
Let us consider the case that $\hat{p}_{\mu}$ consist of $\tilde{d}/2$
canonical pairs $\hat{p}_i,\hat{q}_i$ which satisfy
$[\hat{p}_i,\hat{q}_j]=iB\delta_{ij}$.
We also assume that the  solutions possess
 the discrete symmetry
which exchanges canonical pairs and $\hat{p}_i\leftrightarrow \hat{q}_i$
in each canonical pair.
We then find $V_{p}=\Lambda^{\tilde{d}}$ where $\Lambda$ is the
extension of each
$\hat{p}_{\mu}$.
The volume of the unit quantum in phase space is
$\Lambda^{\tilde{d}}/n=\lambda^{\tilde{d}}$
where $\lambda$ is the spacing of the quanta.
$B$ is related to $\lambda$ as $B=\lambda^2/(2\pi)$.
Let us assume the topology of the world sheet to be
$T^{\tilde{d}}$ in order to determine the distributions of $k_{\mu}$.
Then we can formally construct $\hat{p}_{\mu}$ through unitary matrices
as $\gamma_{\mu}=exp(i2\pi\hat{p}_{\mu}/\Lambda)$.
The polynomials of $\gamma_{\mu}$ are the basis of
$exp(iC^{\mu\nu}k_{\mu}\hat{p}_{\nu})$. We can therefore assume
that $k_{\mu}$ is
quantized in the unit of $|k^{min}|=\lambda/n^{1/ \tilde{d}}$.
The eigenvalues of $\hat{p}_{\mu}$ are quantized in the unit of
$\Lambda/n^{2/ \tilde{d}}=\lambda/n^{1/ \tilde{d}}$.
Hence we restrict the range of $k_{\mu}$ as
$-n^{
1/ \tilde{d}} \lambda/2<k_{\mu}< n^{1/ \tilde{d}}\lambda/2$.
So $\sum_k$ runs over $n^2$ degrees
of freedom which coincide with those of $n$ dimensional Hermitian matrices.

We can construct a map from a matrix to a function
as
\beq
\hat{a} \rightarrow a(x)=\sum_k \tilde{a}(k) exp(ik\cdot x) ,
\label{proj}
\eeq
where $k\cdot x=k_{\mu}x^{\mu}$.
By this construction, we obtain the $\star$ product
\beqa
\hat{a}\hat{b} &\rightarrow& a(x)\star b(x),\n
a(x)\star b(x)&\equiv&exp({C^{\mu\nu}\over 2i}{\partial ^2\over
\partial\xi^{\mu}
\partial\eta^{\nu}})
a(x+\xi )b(x+\eta )|_{\xi=\eta=0} .
\label{star}
\eeqa
The operation $Tr$ over matrices can be exactly mapped onto the integration
over functions as
\beq
Tr[\hat{a}] =
\sqrt{det B}({1\over 2\pi})^{\tilde{d}\over 2}\int d^{\tilde{d}}x a(x) .
\label{traceint}
\eeq
The twisted reduced model can be shown to be equivalent to
noncommutative Yang-Mills by the
the following map from matrices onto functions
\beqa
\hat{a} &\rightarrow& a(x) , \n
\hat{a}\hat{b}&\rightarrow& a(x)\star b(x) ,\n
Tr&\rightarrow&
\sqrt{det B}({1\over 2\pi})^{\tilde{d}\over 2}\int d^{\tilde{d}}x .
\label{momrule}
\eeqa
The following commutator is mapped to the covariant derivative:
\beq
[\hat{p}_{\mu}+\hat{a}_{\mu},\hat{o}]\rightarrow
{1\over i}\partial_{\mu}o(x)+a_{\mu}(x)\star o(x)-o(x)\star a_{\mu}(x)
\equiv [D_{\mu},o(x)] ,
\label{pcovder}
\eeq
We may interpret  the newly emerged coordinate space
as the semiclassical limit of $\hat{x}^{\mu}=C^{\mu\nu}\hat{p}_{\nu}$.
The space-time translation is realized by the following unitary
operator:
\beq
exp(i\hat{p}\cdot d)\hat{x}^{\mu}
exp(-i\hat{p}\cdot d)\n
=\hat{x}^{\mu}+d^{\mu} .
\label{transl}
\eeq

Applying the rule eq.(\ref{momrule}), the bosonic action becomes
\beqa
&&-{1\over 4g^2}Tr[A_{\mu},A_{\nu}][A_{\mu},A_{\nu}]\n
&=&
{\tilde{d}nB^2\over 4g^2}-\sqrt{det B}({1\over 2\pi})^{\tilde{d}\over 2}
\int d^{\tilde{d}}x
{1\over g^2} ({1\over 4}
[D_{\alpha},D_{\beta}][D_{\alpha},D_{\beta}]\n
&&+{1\over 2}[D_{\alpha},
\varphi_{\nu}][D_{\alpha},\varphi_{\nu}]
+{1\over 4}[\varphi_{\nu},\varphi_{\rho}]
[\varphi_{\nu},\varphi_{\rho}])_{\star} .
\eeqa
In this expression, the indices $\alpha,\beta$ run over $\tilde{d}$
dimensional world volume directions  and $\nu,\rho$
over the transverse directions.
We have replaced $a_{\nu}\rightarrow\varphi_{\nu}$ in the transverse
directions. Inside $(~~)_{\star}$, the products should be understood as
$\star$
products and hence commutators do not vanish.
The fermionic action becomes
\beqa
&&{1\over g^2}Tr\bar{\psi}{\Gamma}_{\mu}[A_{\mu},\psi]\n
&=&
\sqrt{det B}({1\over 2\pi})^{\tilde{d}\over 2}
\int d^{\tilde{d}}x{1\over g^2}
(\bar{\psi}{\Gamma}_{\alpha}[D_{\alpha},\psi ]
+\bar{\psi}\Gamma_{\nu}[\varphi_{\nu},\psi ])_{\star} .
\eeqa
We therefore find noncommutative U(1) gauge theory.

In order to obtain noncommutative Yang-Mills theory with $U(m)$ gauge group,
we consider new classical solutions which are obtained by replacing each
element
of $\hat{p}_{\mu}$ by the $m\times m$ unit matrix:
\beq
\hat{p}_{\mu} \rightarrow \hat{p}_{\mu}\otimes {\mathbf{1}}_m .
\eeq
We require $N=mn$ dimensional matrices for this construction.
The fluctuations around this background $\hat{a}$ and $\hat{\psi}$
can be Fourier
decomposed in the analogous way as in eq.(\ref{twist}) with  $m$ dimensional
matrices $\tilde{a}(k)$ and $\tilde{\psi} (k)$ which satisfy
$\tilde{a}(-k)=\tilde{a}^{\dagger}(k)$ and
$\tilde{\psi}(-k)=\tilde{\psi} ^{\dagger}(k)$.
It is then clear that $[\hat{p}_{\mu}+\hat{a}_{\mu},\hat{o}]$ can be mapped
onto the
nonabelian covariant derivative $[D_{\mu},o(x)]$ once we use $\star$ product.
Applying our rule (\ref{momrule}) to the action in this case,
we obtain
\beqa
&&{\tilde{d}NB^2\over 4g^2}-\sqrt{det B}({1\over 2\pi})^{\tilde{d}\over 2}
\int d^{\tilde{d}}x
{1\over g^2} tr({1\over 4}
[D_{\alpha},D_{\beta}][D_{\alpha},D_{\beta}]\n
&&+{1\over 2}[D_{\alpha},\varphi_{\nu}][D_{\alpha},\varphi_{\nu}]
+{1\over 4}[\varphi_{\nu},\varphi_{\rho}][\varphi_{\nu},\varphi_{\rho}]\n
&&+{1\over 2}\bar{\psi}{\Gamma}_
{\alpha}[D_{\alpha},\psi ]
+{1\over 2}\bar{\psi}\Gamma_{\nu}[\varphi_{\nu},\psi ])_{\star} .
\eeqa
where $tr$ denotes taking trace over $m$ dimensional subspace.
The Yang-Mills coupling is found to be $g^2_{NC}=(2\pi )^{\tilde{d}\over
2}g^2/B^{\tilde{d}/2}$.
Therefore it will decrease if the density of quanta in phase space
decreases with fixed $g^2$.

The Hermitian models are invariant under the unitary transformation:
$A_{\mu}\rightarrow UA_{\mu}U^{\dagger},
\psi \rightarrow U\psi U^{\dagger}$. As we shall see, the gauge
symmetry can be embedded in the $U(N)$ symmetry.
We expand $U=exp(i\hat{\lambda} )$ and parameterize
\beq
\hat{\lambda}=\sum_k \tilde{\lambda} (k)
exp(i{k}\cdot\hat{x}) .
\eeq
Under the infinitesimal gauge transformation, we find the fluctuations around
the fixed background transform as
\beqa
\hat{a}_{\mu}&\rightarrow & \hat{a}_{\mu}+i[\hat{p}_{\mu},\hat{\lambda} ]
-i[\hat{a}_{\mu},\hat{\lambda} ] ,\n
\hat{\psi}&\rightarrow & \hat{\psi} -i[\hat{\psi},\hat{\lambda} ] .
\eeqa
We can map these transformations
 onto the gauge transformation
in noncommutative Yang-Mills by our rule eq.(\ref{momrule}):
\beqa
&&a_{\alpha}(x) \rightarrow a_{\alpha}(x) +
{\partial \over \partial x^{\alpha}}\lambda (x)
-ia_{\alpha}(x)\star \lambda (x)+i\lambda (x)\star a_{\alpha}(x) ,\n
&&\varphi_{\nu}(x) \rightarrow \varphi_{\nu}(x)
-i\varphi_{\nu}(x)\star \lambda (x)+i\lambda (x)\star \varphi_{\nu}(x) ,\n
&&\psi (x)\rightarrow \psi (x)
-i\psi (x)\star \lambda (x)+i\lambda (x)\star \psi (x) .
\eeqa

\section{Bi-local field representations}
\setcounter{equation}{0}
\hspace*{\parindent}
We have constructed our mapping rule eq.(\ref{momrule})
by expanding matrices in terms of the wave functions $exp(ik\cdot\hat{x})$.
It is the eigenstate of the adjoint $\hat{P}_{\mu}$ with the
eigenvalue of $k_{\mu}$.
With  the $n \times n$ matrix regularization,
the momenta $k_{\mu}$ are quantized with a unit
$k_{min}= \lambda n^{-1/\tilde{d}}$ and
can take values $|k| < \lambda n^{1/\tilde{d}}/2$.
These momentum eigenstates
correspond to the ordinary plane waves when $|k_{\mu}|<\lambda$.
In noncommutative space-time, it is not possible to consider
states which are localized
in the domain whose volume is smaller than the noncommutative scale.
Therefore if we consider the states with
larger  momentum or smaller longitudinal length scale than the noncommutative
scale,
they must expand in the transverse directions.
Recall that the both momentum space and coordinate
space are embedded in the matrices of twisted reduced models.
They are related by $\hat{x}^{\mu}=C^{\mu\nu}\hat{p}_{\nu}$.
The corresponding eigenstates
such as $exp(ik^1\cdot\hat{x})$ and $exp(ik^2\cdot\hat{x})$
are not commutative to each other if $|k^i_\mu| > \lambda$,
since $exp(ik^1\cdot\hat{x})exp(ik^2\cdot\hat{x})
=exp(ik^2\cdot\hat{x})exp(ik^1\cdot\hat{x})exp(iC^{\mu\nu}k^1_{\mu}
k^2_{\nu})$.
The momentum eigenstate $exp(ik\cdot\hat{x})$ is also written
as $exp(-id\cdot\hat{p})$ where $d^{\mu}=C^{\mu\nu}k_{\nu}$, and
this implies that the eigenstate with $|k_\mu| > \lambda$
is more appropriately interpreted as noncommutative
translation operators (\ref{transl})
rather than the ordinary plane waves.
In other words,
the eigenstate $exp(ik\cdot\hat{x})$ with $|k_{\mu}|>\lambda$
may be interpreted as string like extended objects whose
length is $|C^{\mu\nu}k_{\nu}|$.
\subsection{Operator - bi-local field mapping}
In order to make these statements more transparent, we consider
another representation of matrices in this section.
For simplicity we consider the two dimensional case first:
\beq
[\hat{x}, \hat{y}] = -iC
\eeq
where $C$ is positive.
This  commutation relation is realized by the guiding center coordinates
of  the two dimensional system of electrons
in magnetic field.
The generalizations to arbitrary even $\tilde{d}$ dimensions are
straightforward.
We recall that we have $n$ quanta  with $n$ dimensional matrices.
Each quantum occupies the space-time volume of $2\pi C$.
We may consider a square von Neumann lattice with the lattice
spacing $l_{NC}$ where $l_{NC}^2=2\pi C$.
This spacing $l_{NC}$ gives the noncommutative scale.
Let us denote the most localized state centered at the origin
by $|0 \rangle$.
It is annihilated by the operator $\hat{x}^{-}=\hat{x}-i\hat{y}$.
We construct states localized around each lattice site
by utilizing translation operators
$|x_{i} \rangle =exp(-ix_i \cdot\hat{p})|0 \rangle$.
They are the coherent states on a von Neumann lattice
${\bf x}_i = l_{NC} (n_i {\bf e}^x  + m_i {\bf e}^y)$
where $n,m \in {\bf Z}.$
They are complete but non-orthogonal.
In the case of higher dimensions $\tilde{d}$,
we introduce
\beq
\hat{x}^{\pm a} = \hat{x}^{2a-1} \pm i \hat{x}^{2a}
\eeq
where $a=1,\cdots , \tilde{d}/2$
and define the states $|{\bf x}_i \rangle$ accordingly.
In the following, we set $C^{2a-1, 2a}=C$ for simplicity.
\par
The basic identity in this section is
\beq
\langle 0|exp(ik\cdot\hat{x})|0 \rangle =exp(-{Ck^2\over 4}).
\eeq
Using this identity, we indeed find
\beq
\rho_{ij} \equiv \langle x_i|x_j \rangle =exp({i\over 2}B_{\mu\nu}
x_i^{\mu}x_j^{\nu})exp
(-{(x_i-x_j)^2\over 4 C}).
\eeq
Although  $|x_i \rangle $ are non-orthogonal,
$\langle x_i|x_j \rangle $ exponentially vanishes
when $(x_i-x_j)^2$ gets large.
We note that the following wave functions of the two
dimensional system of free electrons in the lowest Landau level
are exponentially localized around $x_i$:
\beq
c_{n_im_i}(x)={1\over l_{NC}}<x|x_i>.
\eeq

Completeness of the basis leads to the resolution of unity
\beq
1= \sum_{i,j} (\rho^{-1})_{ij} |x_i \rangle \langle x_j|
\eeq
and the trace of an operator $\hat{{\cal O}}$ is given by
\beq
Tr \hat{{\cal O}} = \sum_{i,j} (\rho^{-1})_{ij}
\langle x_j| \hat{{\cal O}}  |x_i \rangle.
\eeq
We also find
\beq
\langle x_i|exp(ik\cdot\hat{x})|x_j \rangle =
exp(ik\cdot{(x_i+x_j)\over 2}+{i\over 2}B_{\mu\nu}x_i^{\mu}x_j^{\nu})
exp(-{(x_i-x_j-d)^2\over 4 C})
\eeq
where $d^{\mu}=C^{\mu\nu}k_{\nu}$.
This matrix element sharply peaks at $x_i-x_j=d$. It supports our
interpretation that
the eigenstate $exp(ik\cdot\hat{x})$ with $|k_{\mu}|>\lambda$
can be interpreted as string like extended objects whose
length is $|C^{\mu\nu}k_{\nu}|$.
When $|k_{\mu}|<\lambda$, on the other hand, this matrix
becomes close to diagonal whose matrix elements go like
\beq
\langle x_i|exp(ik\cdot\hat{x})|x_j \rangle \sim
exp(ik\cdot x_i) \langle x_i|x_j \rangle .
\label{diagonalpart}
\eeq
It again supports our interpretation that
$exp(ik\cdot\hat{x})$ correspond to the ordinary plane waves
when $|k_{\mu}|<\lambda$. They are represented by the
matrices which are close to diagonal.
\par
We now propose the bi-local field representation of noncommutative
field theories.
We may expand matrices $\hat{\phi}$ in the twisted reduced model
by the following bi-local basis as follows:
\beq
\hat{\phi}=\sum_{i,j} \phi(x_i, x_j)|x_i \rangle \langle x_j|
\label{bi-local}
\eeq
where the Hermiticity
 of $\hat{\phi}$ implies $\phi^*(x_j, x_i)=\phi(x_i, x_j)$.
The matrices $\hat{\phi}$ represent $\hat{a_{\mu}}$ or $\hat{\psi}$
in the super Yang-Mills case but the setting
here is more generally applied to
 an arbitrary  noncommutative field theory.
The bi-local basis spans the whole $n^2$ degrees of freedom of matrices.
The product of two operators is also given as
\beq
\hat{\phi}_1 \hat{\phi}_2 = \sum_{i,j,k,l}
 \phi_1(x_i, x_j) \rho_{jk} \phi_2(x_k, x_l) |x_i \rangle \langle x_l|
\eeq
and therefore
\beq
(\phi_1 \phi_2)(x_i, x_j) = \sum_{k,l} \phi_1(x_i, x_k) \rho_{kl}
\phi_2(x_l, x_j).
\eeq
The trace of  operators $\hat{{\cal O}}_1 \cdots \hat{{\cal O}}_s$
in the bi-local basis is
\beq
Tr \hat{{\cal O}}_1 \cdots \hat{{\cal O}}_s
= \sum_{i_1 \cdots i_{2s}} {\cal O}_1 (x_{i_1}, x_{i_2}) \rho_{i_2 i_3}
\cdots  {\cal O}_s (x_{i_{2s-1}}, x_{i_{2s}}) \rho_{i_{2s} i_1}.
\eeq
\par
Here we work out the translation rule between the
momentum eigenstate representation
$\hat{\phi}= \sum_k \tilde\phi(k) exp(ik\cdot\hat{x})$ and the bi-local
field representation of eq.(\ref{bi-local}):
\beqa
 \tilde\phi (k) &=& {1 \over n} Tr (exp(-ik\cdot\hat{x}) \hat{\phi})
= {1 \over n} \sum_{i,j}
\langle x_i| exp(-ik\cdot\hat{x}) |x_j \rangle \
\phi(x_j, x_i) \n
&=& {1 \over n} \sum_{x_c} \phi(x_c) exp(-ik \cdot x_c),\n
\phi(x_c)&=&\sum_{x_r} exp({i\over 2}B_{\mu\nu}x_r^{\mu}x_c^{\nu})
exp(-{(x_r-d)^2\over 4C})\phi (x_j,x_i),
\label{xcxr}
\eeqa
where $x_c=(x_i+x_j)/2$ and $x_r=x_i-x_j$.
From eq.(\ref{xcxr}), we observe that
the slowly varying field with the momentum smaller than $\lambda$
consists of the almost diagonal components.
Hence close to diagonal components of the bi-local field
are identified with the ordinary slowly varying field $\phi(x_c)$.
On the other hand, rapidly oscillating fields are mapped to the
off-diagonal open string states.
A large momentum in $\nu$ direction
$|k_{\nu}| > \lambda $ corresponds to a large distance in the $\mu$-th
direction
$|d^{\mu}| = |C^{\mu \nu} k_{\nu}| > l_{NC}$.
We can decompose $d$ as $d = d_0 + \delta d$
where $d_0$ is a vector which connects two points on the von Neumann lattice
and $|\delta d| < l_{NC}$. Then the summation over $x_r$ in (\ref{xcxr})
is dominated at $x_r = d_0$.
In this way
the large momentum degrees of freedom are more naturally interpreted
as extended open string-like fields.
They are denoted by `open strings' in this paper.
\par
 The adjoint $P^2$ acts on $\hat{\phi}$ as
 \beqa
&& P^2 \hat{\phi}= {B^2\over 2} X^2 \hat{\phi} \n
&=& {B^2\over 2} ( \hat{x}^{+a} \hat{x}^{-a}  \hat{\phi}
- \hat{x}^{-a}  \hat{\phi} \hat{x}^{+a}
- \hat{x}^{+a} \hat{\phi} \hat{x}^{-a}
 + \hat{\phi} \hat{x}^{+a}  \hat{x}^{-a} +  \tilde{d} C \hat{\phi}).
 \eeqa
The kinetic term of bosonic fields is of the form
$- Tr((P_{\mu} \hat{\phi})^2) = Tr(\hat{\phi} P^2 \hat{\phi})$
and  is evaluated in this basis as
\beqa
&&{1\over 2}Tr\hat{\phi}P^2\hat{\phi}=
{B^2\over 2}Tr\hat{\phi}X^2\hat{\phi} \n
&=&
{B^2\over 2} \sum_{i,j,k,l}
\phi(x_i,x_j) \phi(x_k,x_l) \rho_{jk} \rho_{li}
\{ (x_l^{+a}-x_j^{+a})(x_i^{-a}-x_k^{-a})+  \tilde{d} C \},
\label{quadac}
\eeqa
where we have used the property of the coherent basis
$\hat{x}^{-a}|x_i>=x^{-a}_i|x_i>$.

Here we decompose $\rho_{ij}$ by inserting the identity
constructed by the orthogonal basis:
\beq
1=\sum_k|x_k\}\{x_k|.
\eeq
An explicit construction of orthogonal localized wave functions $
w_{m_in_i}(x)=<x|x_i\}$ is carried out
in the two dimensional case\cite{RZE}.
They have a nearly Gaussian shape up to the
radius $r<l_{NC}$ from the center. Outside this region,
it is small, oscillates, and falls off with $r^{-2}$.
The remarkable property of the asymptotic form is that
it vanishes on the von Neumann lattice.
It implies that $<x_i|x_j\}$ vanish for large $|x_i-x_j|$
very rapidly.
Then eq.(\ref{quadac}) can be rewritten for large $|x_i-x_j|$ as
\beqa
&&{B^2\over 2} \sum_{i,j,k,l,n,m}
\phi(x_i,x_j) \phi(x_k,x_l)
<x_j|x_m\}\{x_m|x_k><x_l|x_n\}\{x_n|x_i>\n
&&\times
 (x_l^{+a}-x_j^{+a})(x_i^{-a}-x_k^{-a})\n
&=&
{B^2\over 2} \sum_{n,m}
\check{\phi}(x_n,x_m) \check{\phi}(x_m,x_n)
(x_n-x_m)^2 ,
\eeqa
where
\beq
\check{\phi}(x_i,x_j)=\sum_{k,l}\{x_i|x_k>\phi(x_k,x_l)<x_l|x_j\}=\{x_i|\hat
{\phi}|x_j\}.
\eeq
In this derivation, we have used the fact that $\{x_n|x_i>$ are supported
only for
small $|x_n-x_i|$.

The kinetic term is diagonal for large $|x_i-x_j|$ and we read off
the propagator of the bi-local field
\beq
\langle\check{\phi}(x_i,x_j)\check{\phi}(x_j,x_i)\rangle
={C^2\over (x_i-x_j)^2} .
\label{propagator}
\eeq
Long bi-local fields are thus interpreted to be far off-shell.
It may be reinterpreted as
\beq
 \langle \check{\phi} (p_i,p_j) \check{\phi} (p_j,p_i) \rangle = {1\over
(p_i-p_j)^2},
\eeq
where $p_{i, \mu}=B_{\mu\nu}x_i^{\nu}$.
For small $|x_i-x_j|$, it is more appropriate to use the
momentum eigenstate representation and we can obtain the standard propagator
in that way.
\par
We next consider  three point vertices.
Let us first consider the simplest three point vertex:
\beqa
Tr(\hat{\phi}^3) & = &\sum_{i,j,k,l,m,n} \phi(x_i, x_j) \rho_{jk}
 \phi(x_k, x_l) \rho_{lm} \phi(x_m, x_n) \rho_{ni}\n
&=&\sum_{i,j,k} \check{\phi}(x_i, x_j) \check{\phi}(x_j, x_k)
\check{\phi}(x_k, x_i).
\label{phi3vertex}
\eeqa
We can similarly obtain higher  point vertices as well:
\beqa
Tr (\hat{\phi}^s) &=& \sum_{i_1 \cdots i_{2s}} \phi(x_{i_1}, x_{i_2})
\rho_{i_{2} i_3} \cdots \phi(x_{i_{2s-1}}, x_{i_{2s}})
\rho_{i_{2s} i_1} \n
  &=& \sum_{i_1 \cdots i_{s}} \check{\phi}(x_{i_1}, x_{i_2}) \cdots
\check{\phi}(x_{i_s}, x_{i_1}).
\eeqa
In this way, we make contact with the quenched reduced models
\cite{RM} in the large momentum region.
In quenched reduced models, the eigenvalues of matrices are identified with
momenta.
Therefore we have found that the large momentum behavior of
twisted reduced models is identical to quenched reduced models.
In this correspondence, the relative distance of the two ends
of `open string' in twisted reduced model is related to
the momentum in quenched reduced model by the relation
$\hat{p}_{\mu}=B_{\mu\nu}\hat{x}^{\nu}$.

\subsection{Perturbations}
Perturbative behaviors of noncommutative field theories have been
investigated \cite{Jab,KW,MS,ChRo,MRS,ABK,hayakawa}.
In particular, it was pointed out
that the effective action exhibits
infrared singular
behaviors due to the nonplanar diagrams\cite{MRS,hayakawa}.
We now look at the perturbation of the noncommutative field theories
in the bi-local basis.
Let us consider the $\phi^3$ theory as a simple example.
The matrix  model action is given by
\beq
S = Tr \left(  -{1\over 2} [\hat{p}_{\mu},\hat{\phi}]^2
+ {\lambda\over 3} \hat{\phi}^3 \right) .
\label{phi3action1}
\eeq
In the large momentum region, the propagator and vertex are
 given by (\ref{propagator})  and (\ref{phi3vertex}).
In the one-loop approximation, there are two types of diagrams.
\par
\begin{figure}[h]
\begin{center}
\leavevmode
\epsfxsize=6cm
\epsfbox{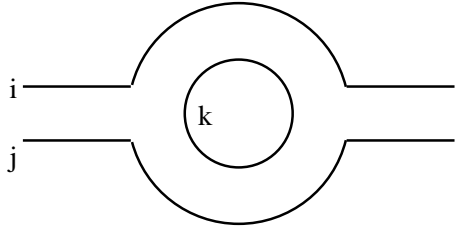}
\caption{A planar graph at one loop: it renormalizes the
  kinetic term of bi-local fields $\check{\phi}(x_i, x_j)$.}
\label{fig:planar}
\end{center}
\end{figure}

Fig. \ref{fig:planar} is planar and gives the correction to the propagator
of the bi-local field $\check{\phi}(x_i,x_j)$:
\beq
\sum_{i,j} \sum_k {C^4 \check{\phi}(x_i,x_j) \check{\phi}(x_j,x_i)
\over (x_i-x_k)^2 (x_j-x_k)^2 }.
\label{planar1}
\eeq
This diagram corresponds to the  following planar diagram amplitude
in the conventional perturbative expansion of
noncommutative field theories:
\beq
\sum_{p,k} {\tilde{\phi}(p) \tilde{\phi}(-p) \over k^2 (p-k)^2}.
\label{planar2}
\eeq
As originally proved in \cite{twisted}, noncommutative phases
which can be assigned to the propagators are canceled in planar diagrams
and the amplitudes of such graphs are the same as the commutative cases
except for the phase factors which only depend on the external momenta.
In $\tilde{d}$-dimensions, the integral (\ref{planar2}) is UV divergent
as $\Lambda^{\tilde{d}-4}$ where $\Lambda$ is the ultraviolet cutoff.
In the bi-local representation (\ref{planar1}),
the summation is also divergent but it originates from the infrared
extension of the von Neumann lattice.
It is reminiscent of the tadpole divergences in string theory.
However this analogy requires more work to substantiate it.
For example,  we may be able to work out a direct
world sheet interpretation of the bi-local propagators.
\par
\begin{figure}[h]
\begin{center}
\leavevmode
\epsfxsize=6cm
\epsfbox{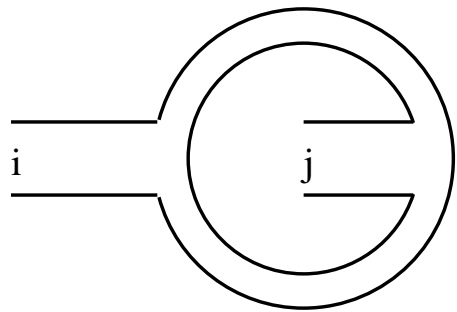}
\caption{A nonplanar graph at one loop: it
generates long range interaction between diagonal components of
   the bi-local field $\check{\phi}(x_i, x_i)$.
}
\label{fig:nonplanar}
\end{center}
\end{figure}

Fig. \ref{fig:nonplanar}
 is a nonplanar diagram and induces effective interactions
for only the
diagonal components $\check{\phi}(x_i, x_i)$ of the bi-local fields.
Since the diagonal components are interpreted as
ordinary (slowly varying ) local fields $\phi(x_i)$,
this diagram induces long range interactions:
\beq
S_{eff} = C^4 \sum_{i,j} {{\phi}(x_i) {\phi}(x_j) \over (x_i-x_j) ^4}.
\label{nonplanar1}
\eeq
This type of interactions have been well known
as the block-block interactions in the
matrix models \cite{IKKT}.
In the conventional picture of noncommutative $\phi^3$ theory,
this diagram provides the following  nonplanar one-loop correction
to the propagator.
\beq
\sum_{p,k} {\tilde{\phi}(p) \tilde{\phi}(-p)
 \over k^2 (p-k)^2} exp(i C_{\mu \nu}k^{\mu} p^{\nu}).
\label{nonplanar2}
\eeq
It is very specific to noncommutative field theory since
it exhibits infrared singular behaviors for small $p$.

A physically intuitive argument which relates
the nonplanar amplitude in the bi-local basis (\ref{nonplanar1}) and the
conventional
nonplanar amplitude (\ref{nonplanar2}) is given as follows:
\beqa
S_{eff} &=& C^4 \sum_{p,q} \sum_{i,j}
{\tilde{\phi}(p) \tilde{\phi}(q) \over (x_i-x_j) ^4}
e^{ip \cdot x_i} e^{iq \cdot x_j} \n
&=& n C^4 \sum_p \sum_d {\tilde{\phi}(p) \tilde{\phi}(-p) \over d ^4}
e^{ip \cdot d}.
\eeqa
If we rewrite $k_{\mu}=B_{\mu \nu} d^{\nu}$, this has the same expression
as eq.(\ref{nonplanar2}). Here we also recall that the momentum lattice
$\{k_{\mu}\}$ has finer
resolutions than the von Neumann lattice $\{d_{\mu}\}$. Therefore
$\sum_k=n\sum_d$
for large $|d|$. Note that the nonplanar phase is interpreted as
the wave functions of plane waves here and
the summation over $k$ in (\ref{nonplanar2}) amounts to performing the Fourier
transformation. On the other hand,
the summation over $k$ in (\ref{planar2})
leads to the standard one loop integration  in the planar contribution.
\par
From these arguments, it is clear that the long range interaction
$1/d^4$  is still induced even when the $\phi$ field has a mass term $m$
as long as $d > Cm$.
In the next section, we investigate these long range interactions
of matrix models in more detail.
\section{Interactions between diagonal blocks}
\setcounter{equation}{0}
\hspace*{\parindent}
In this section, we investigate the renormalization property of
noncommutative field theory.
After integrating long `open strings', we obtain long range
interactions. This phenomena appear as nonplanar contributions
in noncommutative field theory\cite{MRS}.
We show that it can be simply understood as block-block
interactions in the matrix model picture.
\subsection{Noncommutative $\phi^3$}
We consider a noncommutative
$\phi^3$ field theory as a simple example.
The matrix  model action is given by (\ref{phi3action1}).
Following the same procedure as in eqs. (\ref{proj} - \ref{momrule}),
we can obtain a noncommutative $U(m)$ $\phi^3$ field theory:
\beq
S = \int d^{\tilde{d}}x \ tr \left(
{1\over 2} (\partial_{\mu} \phi(x))^2 + {\lambda '\over 3} \phi(x)^3
\right)_{\star}.
\eeq
Here $tr$ means a trace over $(m \times m)$ matrices and
$\lambda '=\lambda ({2\pi / B})^{\tilde{d}/ 4}$.
\par
We consider a group of backgrounds $\phi^{(i)}$. We assume that
the corresponding functions $\phi^{(i)} (x)$ are localized
and well separated.
$\phi^{(i)}$ can be specified by its Fourier components
$\tilde{\phi}^{(i)} (k)$:
\beq
\hat{\phi}^{(i)}=\sum_k \tilde{\phi}^{(i)} (k)exp(ik\cdot \hat{x}).
\eeq
We further assume that
$\tilde{\phi}^{(i)} (k)$ is only supported for $|k|<< \lambda$ which implies
the much larger localization scale than the noncommutativity scale.
This condition means that the background
in the bi-local representation $ \phi^{(i)} (x, y)$ is almost
diagonal.
We hence expect that the following representation is
accurate :
\beq
\hat{\phi}^{(i)} = \sum_j \phi^{(i)} (x_j) |x_j \rangle \langle x_j|.
\eeq
The background is assumed to be localized around $x = d^{(i)}$.
In this section we obtain the effective action for these
backgrounds by integrating over the off-diagonal components of
the bi-local field, that is, the higher momentum fields.
\par
The  assumption of no overlap
imposes certain conditions on $\tilde{\phi}^{(i)} (k)$.
The commutators are mapped in our mapping rule as follows:
\beq
[\phi^{(i)},\phi^{(j)}]\rightarrow
\phi^{(i)}(x)\star\phi^{(j)}(x)-\phi^{(j)}(x)\star\phi^{(i)}(x).
\label{comrel}
\eeq
Then we can see that $\phi^{(i)}$ and $\phi^{(j)}$ are commutative
to each other from eq.(\ref{comrel}). Therefore they can be represented by
a block diagonal form as follows:
\beq
\hat{\phi}_{cl}= \sum \hat{\phi}^{(i)}=
\left(\begin{array}{cccc}
                    \phi^{(1)}  &                &               &   \\
                                   & \phi^{(2)}  &               &   \\
                                   &                & \phi^{(3)} &   \\
                                   &                &               &  \ddots
                 \end{array} \right) ,
\label{blockdiagonal}
\eeq
where $\phi^{(i)}$ $(i=1, 2, \cdots)$ is a $n_i\times n_i$ matrix.

$[\hat{p}_{\mu},\phi^{(i)}]$ can be mapped onto $-i{\partial}_{\mu}
\phi^{(i)}(x)$.
Since ${\partial}_{\mu}\phi^{(i)}(x)\star\phi^{(j)}(x)-
\phi^{(j)}(x)\star{\partial}_{\mu}\phi^{(i)}(x)$ vanishes for well
separated
wave-packets, $[\hat{p}_{\mu},\phi^{(i)}]$ are also of the block diagonal form
as in eq.(\ref{blockdiagonal}). This fact in turn implies that $\hat{p}_{\mu}$
can be represented in the same block diagonal form when it acts on these
backgrounds:
\beq
\hat{p}_{\mu}=\left(\begin{array}{cccc}
                    p_{\mu}^{(1)}  &                &               &   \\

                                   & p_{\mu}^{(2)}  &               &   \\
                                   &                & p_{\mu}^{(3)} &   \\
                                   &                &               &  \ddots
                 \end{array} \right) ,
\label{blockp}
\eeq
Here $p_{\mu}^{(i)}$ denotes the projection of $\hat{p}_{\mu}$ onto
the subspace specified by $\phi^{(i)}$. $\hat{p}_{\mu}$ can be represented
as the block diagonal form since we have projected it onto the subspace
where the backgrounds $\phi^{(i)}$ are supported.
\par
We now calculate the one-loop effective action between
diagonal blocks
in order
 to investigate the renormalization property of noncommutative
field theory.
We may further decompose ${p}_{\mu}^{(i)}$ as
\beqa
{p}_{\mu}^{(i)}&=&B_{\mu\nu}d_{\nu}^{(i)}1_{n_i}+\tilde{p}_{\mu}^{(i)} ,\n
Tr\tilde{p}_{\mu}^{(i)}&=& 0 ,
\label{cm}
\eeqa
where $d_{\nu}^{(i)}$ is a $c$ number representing the center of mass
coordinate of the $i$-th block.
Here we assume that the blocks are separated far enough from each other,
that is,
for all $i$ and $j$'s, $(d_{\mu}^{(i)}-d_{\mu}^{(j)})^2$'s are larger
than the localization scale of each block.

We first recall some notations which are introduced in \cite{IKKT}.
We denote the $(i,j)$ block of a matrix $X$ as $X^{(i,j)}$. It is clear that
$P_{\mu}=[\hat{p}_{\mu},~]$ operates on each $X^{(i,j)}$
independently. In fact we have
\beq
(P_{\mu}X)^{(i,j)}=B_{\mu\nu}(d_{\nu}^{(i)}-d_{\nu}^{(j)}) X^{(i,j)}
                      +\tilde{p}_{\mu}^{(i)}X^{(i,j)}
                                    -X^{(i,j)}\tilde{p}_{\mu}^{(j
)}.
\eeq
In the bi-local basis developed in the previous section,
$(i,j)$ block of a matrix $X^{(i,j)}$ may correspond to a
collections of bi-local
fields which connect the $i$-th and $j$-th diagonal blocks.
We further simplify this equation by introducing notations such as
\beqa
d_{\mu}^{(i,j)}X^{(i,j)}&=&(d_{\mu}^{(i)}-d_{\mu}^{(j)}) X^{(i,j)} ,\n
P_{L\mu}^{(i,j)}X^{(i,j)}&=&\tilde{p}_{\mu}^{(i)}X^{(i,j)} ,\n
P_{R\mu}^{(i,j)}X^{(i,j)}&=&-X^{(i,j)}\tilde{p}_{\mu}^{(j)} .
\eeqa
Note that $d_{\mu}^{(i,j)}$,$P_{L\mu}^{(i,j)}$ and $P_{R\mu}^{(i,j)}$
commute each other, and
the operation of $P_{\mu}$ on $X^{(i,j)}$ is expressed as
\beq
(P_{\mu}X)^{(i,j)}=(B_{\mu\nu}d_{\nu}^{(i,j)}+P_{L\mu}^{(i,j)}
                              +P_{R\mu}^{(i,j)})X^{(i,j)}.
\label{lrdecomposition}
\eeq
We can also decompose the action of ${\hat{\phi}}_{cl}$ onto $X^{(i,j)}$,
which should be made symmetric between the right and the left multiplications
for Hermiticity,
in the same way as (\ref{lrdecomposition}):
\beq
(\Phi X)^{(i,j)} \equiv (\Phi_{L}^{(i,j)}
                          +\Phi_{R}^{(i,j)})X^{(i,j)},
\label{lrdecompositionF}
\eeq
where
\beqa
\Phi_{L}^{(i,j)}X^{(i,j)}&=& \phi^{(i)}X^{(i,j)} ,\n
\Phi_{R}^{(i,j)}X^{(i,j)}&=&X^{(i,j)}\phi^{(j)}.
\eeqa
Since the left and right multiplication are totally independent, we have
\beqa
(OX)^{(i,j)}&\equiv&O^{(i,j)}_LX^{(i,j)}O^{(i,j)}_R,\n
{\cal T}r  \, O &=&\sum_{i,j=1}^{n} Tr O_L^{
(i,j)} Tr O_R^{(i,j)},
\label{ijtrace}
\eeqa
for operators consisting of $P_{\mu}$ and $\Phi$.
Here ${\cal T}r$ denotes the trace of the operators
which act on the matrices.
\par
The one loop effective action is
\beq
W={1\over 2}{\cal T}rlog(1-{1\over P_{\mu}^2}\lambda\Phi)).
\eeq
Now we expand the expression of the one-loop effective action
with respect to the inverse power of $d_{\mu}^{(i,j)}$'s.
We drop the linear term in the background fields since we adopt
the background field method which has no ambiguity in this case
unlike the gauge fixing ambiguity in gauge theory.
We have,
\beqa
W    &=&-{1\over 4}{\cal T}r
    \left(\frac{1}{P^2}\lambda\Phi \frac{1}{P^2}\lambda\Phi
              \right) \n
    &~& -{1\over 6}{\cal T}r
    \left(\frac{1}{P^2}\lambda\Phi \frac{1}{P^2}\lambda\Phi
             \frac{1}{P^2}\lambda\Phi\right) \n
    &~&-\frac{1}{8} {\cal T}r
    \left(\frac{1}{P^2}\lambda\Phi \frac{1}{P^2}\lambda\Phi
             \frac{1}{P^2}\lambda\Phi\frac{1}{P^2}\lambda\Phi \right)
       +O((\Phi)^5).
\label{Wexpansion}
\eeqa
Since as in (\ref{lrdecomposition}) and (\ref{lrdecompositionF}) $P_{\mu}$
and $\Phi$ act
on the $(i,j)$ blocks independently, the one-loop effective
action $W$ is expressed as the sum of contributions of the $(i,j)$ blocks
$W^{(i,j)}$. Therefore we may consider $W^{(i,j)}$
as the interaction between the $i$-th and $j$-th blocks. Using (\ref{ijtrace})
and (\ref{Wexpansion}) we can easily evaluate $W^{(i,j)}$
to the leading order of $1/\sqrt{(d^{(i)}-d^{(j)})^2}$ as
\beqa
W^{(i,j)}&=&\frac{C^4}{(d^{(i)}-d^{(j)})^4}(-{\lambda^2\over 4})
                    {\cal T}r^{(i,j)}\left(\Phi \Phi \right) \n
       &&+ O((1/(d^{(i)}-d^{(j)})^5)\n
&=&\frac{C^4}{(d^{(i)}-d^{(j)})^4}(-{\lambda^2\over 4})(n_jTr(\phi^{(i)}
\phi^{(i)})
+n_iTr(\phi^{(j)}\phi^{(j)})+2Tr(\phi^{(i)})Tr(\phi^{(j)}))\n
&&+ O((1/(d^{(i)}-d^{(j)})^5).
\label{blockinteraction}
\eeqa
The third term in the above expression can be interpreted as the
signature for the existence of
massless particles corresponding to a scalar in six dimensions.
\par
A more conventional approach to calculate ${\cal T}r$ is to use the
plane-wave basis
$exp(ik\cdot \hat{x})$.
It corresponds to a standard one loop calculation
in noncommutative $\phi^3$:
\beqa
W   &=& -{1\over 4}{ \cal T}r
    \left(\frac{1}{P^2}\lambda\Phi \frac{1}{P^2}\lambda\Phi
              \right) \n
&=&-{1\over 4n}\sum_k Tr(exp(-ik\cdot \hat{x})
\frac{1}{P^2}\lambda\Phi \frac{1}{P^2}\lambda\Phi exp(ik\cdot \hat{x}) )\n
&=&-{\lambda^2\over 4}\sum_l\tilde{\phi}_{cl} (-l)\tilde{\phi}_{cl} (l)
\sum_k {2\over k^2 (k+l)^2}
(1+exp(iC^{\mu\nu}k_{\mu}l_{\nu})).
\eeqa
Here we find the both planar and nonplanar contributions.
The latter contains the nontrivial phase factor $exp(iC^{\mu\nu}
k_{\mu}l_{\nu})$.

We evaluate the nonplanar contributions:
\beqa
&&{1\over n}\sum_k {1\over k^2 (k+l)^2}
exp(iC^{\mu\nu}k_{\mu}l_{\nu})\n
&=&({1\over 2\pi B})^{\tilde{d}\over 2}
\int d^{\tilde{d}}k {1\over k^2 (k+l)^2}
exp(iC^{\mu\nu}k_{\mu}l_{\nu})\n
&=&({1\over 2\pi B})^{\tilde{d}\over 2}\int d\alpha_1d\alpha_2
({\pi\over \alpha_1+\alpha_2})^{\tilde{d}\over 2}
exp(-{\alpha_1\alpha_2\over \alpha_1+\alpha_2}l^2-{(Cl)^2\over
4(\alpha_1
+\alpha_2)}).
\eeqa
When $\tilde{d}>4$, the above integral is evaluated as
\beq
\Gamma ({\tilde{d}\over 2}-2)({C\over 2})^{4-{\tilde{d}\over 2}}
(l^2)^{2-{\tilde{d}\over 2}}.
\eeq
We therefore find the nonplanar contribution:
\beqa
&&-n{\lambda^2\over 2}\sum_l\tilde{\phi} (-l)\tilde{\phi} (l)
({C\over 2})^{4-{\tilde{d}\over 2}}(l^2)^{2-{\tilde{d}\over 2}}
\Gamma ({\tilde{d}\over 2}-2)\n
&=&-n^2{\lambda^2C^4\over 2}({1\over 2})^{4-{\tilde{d}\over 2}}
\Gamma ({\tilde{d}\over 2}-2)({1\over 2\pi })^{\tilde{d}\over 2}
\int d^{\tilde{d}}l\tilde{\phi} (-l)\tilde{\phi} (l)
(l^2)^{2-{\tilde{d}\over 2}}\n
&=&-{\lambda^2C^4\over 2}({1\over 2\pi C})^{\tilde{d}}
\int d^{\tilde{d}}x\int d^{\tilde{d}}y\phi (x){1\over (x-y)^4}\phi (y)
\label{nonplain}
\eeqa
where we have used the identity:
\beq
\int {d^{\tilde{d}}k \over (2\pi )^{\tilde{d}}}exp(ik\cdot x)
({1\over k^2})^{{\tilde{d}\over 2}-2}=
{1\over 2^{\tilde{d}-4}}({1\over \pi})^{\tilde{d}\over 2}
{1\over \Gamma ({\tilde{d}\over 2}-2)}{1\over |x|^4}.
\eeq
We also note that
our convention is
\beq
\tilde{\phi}_{cl} (l)=
{1\over n(2\pi C)^{\tilde{d}\over 2}}\int d^{\tilde{d}}x
exp(-il\cdot x)\phi_{cl} (x).
\eeq

We observe that eq.(\ref{nonplain}) can be identified with the last term of
eq.(\ref{blockinteraction}) since it can be reexpressed as follows
due to our mapping rule eq.(\ref{momrule}):
\beqa
&&\sum_{ij}
\frac{C^4}{(d^{(i)}-d^{(j)})^4}(-{\lambda^2\over 2})Tr(\phi^{(i)})
Tr(\phi^{(j)})\n
&=&-{\lambda^2C^4\over 2}({1\over 2\pi C})^{\tilde{d}}\int d^{\tilde{d}}x
\int d^{\tilde{d}}y
tr \phi_{cl} (x){1\over (x-y)^4}tr \phi_{cl} (y).
\eeqa
Here we have replaced the trace over the
localized classical background $\phi_{i}$ by
the integration over the space-time coordinates in the neighborhood
of the wave packet. By integrating over the entire space-time, we
also sum over  different localized blocks.
Since we are only considering the interactions of the well separated
backgrounds, the above formula is only valid in the low momentum region.
What we have found here is that the nonplanar contributions which give rise
to the long range interactions can be identified with those from off-diagonal
components in matrix models. We can simply reproduce it by considering
block-block interactions. These are distinguished features of noncommutative
field
theory which are absent in field theory.
\subsection{Noncommutative $\phi ^4$ }
We consider noncommutative $\phi ^4$ theory next
\beq
S = Tr \left(  -{1\over 2} [\hat{p}_{\mu},\hat{\phi}]^2
+ {\lambda\over 4} \hat{\phi}^4 \right) .
\eeq
The one loop effective action is
\beq
W={1\over 2} {\cal T}rlog(1-{1\over P_{\mu}^2}
\lambda(\Phi_L^2+\Phi_R^2+\Phi_L\Phi_R)).
\eeq
Now we expand the expression of the one-loop effective action
with respect to the inverse power of $d_{\mu}^{(i,j)}$'s.
We have,
\beqa
W    &=&-{1\over 2}{\cal T}r
\left(\frac{1}{P^2}\lambda(\Phi_L^2+\Phi_R^2+\Phi_L\Phi_R)
              \right)\n
     &&-{1\over 4}{\cal T}r
      \left(\frac{1}{P^2}\lambda(\Phi_L^2+\Phi_R^2
+\Phi_L\Phi_R)
 \frac{1}{P^2}\lambda(\Phi_L^2+\Phi_R^2+\Phi_L\Phi_R)
              \right)
       +O((\Phi)^5).
\eeqa
We can easily evaluate $W^{(i,j)}$
to the leading order of $1/\sqrt{(d^{(i)}-d^{(j)})^2}$ as
\beqa
W^{(i,j)}&=&\frac{1}{(d^{(i)}-d^{(j)})^2}(-{\lambda\over 2})
{\cal T}r^{(i,j)}\left(\Phi_L^2+\Phi_R^2+\Phi_L\Phi_R\right) \n
       &&+ O((1/(d^{(i)}-d{(j)})^3)\n
&=&\frac{1}{(d^{(i)}-d^{(j)})^2}(-{\lambda \over 2})(n_jTr(\phi^{(i)}
\phi^{(i)})
+n_iTr(\phi^{(j)}\phi^{(j)})+Tr(\phi^{(i)})Tr(\phi^{(j)}))\n
&&+ O((1/(d^{(i)}-d^{(j)})^3).
\label{blockint}
\eeqa
We find
the exchanges of massless particles
corresponding to a scalar in four dimensions.
\par
We also evaluate the leading term of the effective action by using the
plane-wave basis
$exp(ik\cdot \hat{x})$.
\beqa
W    &=&-{1\over 2} {\cal T}r
   \left(\frac{1}{P^2}\lambda(\Phi_L^2+\Phi_R^2+\Phi_L\Phi_R)
              \right)\n
&=&-{1\over 2n}\sum_k Tr(exp(-ik\cdot \hat{x})
 \frac{1}{P^2}\lambda(\Phi_L^2+\Phi_R^2+\Phi_L\Phi_R)exp(ik\cdot \hat{x}) )\n
&=&-{\lambda\over 2}\sum_l\tilde{\phi} (-l)\tilde{\phi} (l)\sum_k {1\over k^2 }
(2+exp(iC^{\mu\nu}k_{\mu}l_{\nu})).
\eeqa
Here we also find the both planar and nonplanar contributions.
\par
We evaluate the nonplanar contributions:
\beqa
&&{1\over n}\sum_k {1\over k^2 }
exp(iC^{\mu\nu}k_{\mu}l_{\nu})\n
&=&({1\over 2\pi B})^{\tilde{d}\over 2}
\int d^{\tilde{d}}k {1\over k^2 }
exp(iC^{\mu\nu}k_{\mu}l_{\nu})\n
&=&({1\over 2\pi B})^{\tilde{d}\over 2}\int d\alpha
({\pi\over \alpha})^{\tilde{d}\over 2}
exp(-{(Cl)^2\over 4\alpha}).
\eeqa
When $\tilde{d}>2$, the above integral is evaluated as
\beq
\Gamma ({\tilde{d}\over 2}-1)({C\over 2})^{2-{\tilde{d}\over 2}}
(l^2)^{1-{\tilde{d}\over 2}}.
\eeq
We therefore find the nonplanar contribution:
\beqa
&&-n{\lambda\over 2}\sum_l\tilde{\phi} (-l)\tilde{\phi} (l)
({C\over 2})^{2-{\tilde{d}\over 2}}(l^2)^{1-{\tilde{d}\over 2}}
\Gamma ({\tilde{d}\over 2}-1)\n
&=&-n^2
{\lambda C^2\over 2}({1\over 2})^{2-{\tilde{d}\over 2}}\Gamma
({\tilde{d}\over 2}-1)
({1\over 2\pi })^{\tilde{d}\over 2}
\int d^{\tilde{d}}l\tilde{\phi} (-l)\tilde{\phi} (l)
(l^2)^{1-{\tilde{d}\over 2}}\n
&=&-{\lambda C^2\over 2}({1\over 2\pi C})^{\tilde{d}}
\int d^{\tilde{d}}x\int d^{\tilde{d}}y\phi (x){1\over (x-y)^2}\phi (y)
\label{nonplain2}
\eeqa
where we have used the identity:
\beq
\int {d^{\tilde{d}}k \over (2\pi )^{\tilde{d}}}exp(ik\cdot x)
({1\over k^2})^{{\tilde{d}\over 2}-1}=
{1\over 2^{\tilde{d}-2}}({1\over \pi})^{\tilde{d}\over 2}
{1\over \Gamma ({\tilde{d}\over 2}-1)}{1\over |x|^2}.
\eeq
We observe that eq.(\ref{nonplain2}) can be identified with the last term of
eq.(\ref{blockint}) since it can be reexpressed as follows
due to our mapping rule:
\beqa
&&\sum_{ij}
\frac{C^2}{(d^{(i)}-d^{(j)})^2}(-{\lambda\over 2})Tr(\phi^{(i)})
Tr(\phi^{(j)})\n
&=&-{\lambda C^2\over 2}({1\over 2\pi C})^{\tilde{d}}\int d^{\tilde{d}}x\int
d^{\tilde{d}}y
tr \phi_{cl} (x){1\over (x-y)^2} tr \phi_{cl} (y).
\eeqa
\section{Block-block interactions in gauge theory}
\setcounter{equation}{0}
In the previous section we have seen that the long range
block-block interactions are induced in the effective action for
the slowly varying fields of noncommutative scalar theories
after integrating out the off-diagonal components.
They correspond to nonplanar diagrams which carry nontrivial
phase factors.

These are the familiar interactions in matrix models
and indeed in IIB matrix model they are interpreted
as massless particle propagations \cite{IKKT}.
There is a conceptual difference, however, between scalar theory
and gauge theory.
In scalar theory we must introduce extra matrices
$\hat{p}_{\mu}$ to represent the noncommutative space-time.
In reduced models of gauge theory, they are considered as special
backgrounds of the dynamical variables
($A_{\mu}=\hat{p}_{\mu}+\hat{a}_{\mu}$).
The block-block interactions in reduced models of gauge theory such as
type IIB matrix model are universal and not restricted
to specific backgrounds $\hat{p}_{\mu}$ which define the twisted
reduced models.
Actually the original calculation of such interactions
in \cite{IKKT} does not assume any specific conditions for
backgrounds.
We have obtained the long range interactions
which  decay as $1/r^8$.
It does not depend whether we assume uniform distributions
of the matrix eigenvalues in ten dimensions or
in lower dimensions such as four as is discussed in this section.
These long range interactions are interpreted as propagations of the massless
type IIB supergravity multiplets and this fact can be
considered as  one of the evidences
that IIB matrix model contains gravity.
\par
In this section we reexamine the  block-block
interactions of D-instantons in IIB matrix model\cite{AIIKKT}.
There we have considered the backgrounds which represent four dimensional
noncommutative
space-time. By expanding IIB matrix model around such backgrounds, we obtain
four dimensional noncommutative Yang-Mills theory. There are
nontrivial classical solutions in this
theory which reduce to instantons in the large instanton size limit\cite{NS}.
\par
We have considered a classical solution of IIB matrix model
which represents an instanton and an (anti)instanton.
We can realize $U(4)$ gauge theory by considering four D3 branes.
We embed an instanton into the first $SU(2)$ part and the other
(anti)instanton into the remaining $SU(2)$ part. We separate
them in the fifth dimension by the distance $b$:
\beqa
A_0&=&\left( \begin{array}{cc}
p_0+a_0 & 0 \\
0 & p_0+a'_0
\end{array}
\right) \n
A_1&=&\left( \begin{array}{cc}
p_1+a_1 & 0 \\
0 & p_1+a'_1
\end{array}
\right) \n
A_2&=&\left( \begin{array}{cc}
p_2+a_2 & 0 \\
0 & p_2+a'_2
\end{array}
\right) \n
A_3&=&\left( \begin{array}{cc}
p_3+a_3 & 0 \\
0 & p_3+a'_3
\end{array}
\right) \n
A_4&=&\left( \begin{array}{cc}
{b\over 2} & 0 \\
0 & -{b\over 2}
\end{array}
\right) \n
A_{\rho}&=&0
\eeqa
where $\rho=5,\cdots ,9$.

While two instanton system receives no quantum corrections,
the instanton - anti-instanton system
receives quantum corrections since it is no longer BPS.
We have evaluated the one loop effective potential due to an instanton
and (anti)instanton.
They are local excitations and couple to gravity.
These solutions are characterized by the adjoint field strength $F_{\mu\nu}$
which does not vanish at the locations of the instantons.
We assume that they are separated by a long distance compared
to their sizes.
We also assume that $b >> l_{NC}$.
Then we can choose two disjoint blocks
in each of which a large part of an (anti)instanton is contained.
Let the location and the size of the two instantons $(x_i,\rho_i)$
and  $(x_j,\rho_j )$.\footnote{Here we have used the indices $i$ and $j$ to
denote two
instantons since they are also represented by the diagonal blocks in the
matrix model.
However we consider only two instantons in this section.}
The ten dimensional distance of them is $r^2=(x_i-x_j)^2+b^2$.
Here we have assumed that $r >> \rho$.

The one-loop effective action
of IIB matrix model is
\beq
ReW  = \frac{1}{2}{\cal T}r \log(P_{\lambda}^2 \delta_{\mu\nu}-2iF_{\mu\nu})
      -\frac{1}{4}{\cal T}r
\log((P_{\lambda}^2+\frac{i}{2}F_{\mu\nu}\Gamma^{\mu\nu})
      (\frac{1+\Gamma_{11}}{2}))-{\cal T}r \log(P_{\lambda}^2).
\label{oneloopeffpot}
\eeq
Here $P_{\mu}$ and $F_{\mu\nu}$ are operators acting on the space of
matrices as
\beqa
        P_{\mu}X & = & [p_{\mu},X] ,\n
        F_{\mu\nu}X & = & \left[ f_{\mu\nu},X \right],
\label{adjointoperator}
\eeqa
where $f_{\mu \nu}=i[p_{\mu},p_{\nu}]$.
Now we expand the general expression of the one-loop effective action
(\ref{oneloopeffpot}) with respect to the inverse power of $d_{\mu}^{(i,j)}$'s
just like the preceding section.
We can take traces of the $\gamma$ matrices after expanding the logarithm in
(\ref{oneloopeffpot}).
Due to the supersymmetry, contributions of bosons and fermions cancel
each other to the third order in $F_{\mu\nu}$, and we have,
\beqa
W    &=&-{\cal T}r
   \left(\frac{1}{P^2}F_{\mu\nu} \frac{1}{P^2}F_{\nu\lambda}
             \frac{1}{P^2}F_{\lambda\rho}\frac{1}{P^2}F_{\rho\mu} \right) \n
       &~& -2{\cal T}r
     \left(\frac{1}{P^2}F_{\mu\nu} \frac{1}{P^2}F_{\lambda\rho}
             \frac{1}{P^2}F_{\mu\rho}\frac{1}{P^2}F_{\lambda\nu} \right) \n
       &~&+\frac{1}{2}{\cal T}r
   \left(\frac{1}{P^2}F_{\mu\nu} \frac{1}{P^2}F_{\mu\nu}
             \frac{1}{P^2}F_{\lambda\rho}\frac{1}{P^2}F_{\lambda\rho} \right)
\n
      &~&+\frac{1}{4}{\cal T}r
  \left(\frac{1}{P^2}F_{\mu\nu} \frac{1}{P^2} F_{\lambda\rho}
             \frac{1}{P^2}F_{\mu\nu}\frac{1}
{P^2}F_{\lambda\rho} \right)
       +O((F_{\mu\nu})^5).
\label{Wexpansion2}
\eeqa
Since as in (\ref{lrdecomposition}) and (\ref{lrdecompositionF}) $P_{\mu}$
and $F_{\mu\nu}$ act on the $(i,j)$ blocks independently, the one-loop
effective
action $W$ is expressed as the sum of contributions of the $(i,j)$ blocks
$W^{(i,j)}$. Therefore we may consider $W^{(i,j)}$
as the interaction between the $i$-th and $j$-th blocks.

Using (\ref{ijtrace})
and (\ref{Wexpansion2}) we can easily evaluate $W^{(i,j)}$
to the leading order of $1/\sqrt{(d^{(i)}-d^{(j)})^2}$ as
\beqa
W^{(i,j)}&=&
{1\over r^8}
(-{\cal T}r^{(i,j)}
(F_{\mu\nu}F_{\nu\lambda}F_{\lambda\rho}F_{\rho\mu})
-2{\cal T}r^{(i,j)}
(F_{\mu\nu}F_{\lambda\rho}F_{\mu\rho}F_{\lambda\nu})\n
&&+{1\over 2}{\cal T}r^{(i,j)}
(F_{\mu\nu}F_{\mu\nu}F_{\lambda\rho}F_{\lambda\rho})
+{1\over 4}{\cal T}r^{(i,j)}
(F_{\mu\nu}F_{\lambda\rho}F_{\mu\nu}F_{\lambda\rho})
\n
&=&{3\over 2 r^8}
( -n_j \tilde{b}_8(f^{(i)})- n_i \tilde{b}_8(f^{(j)}) \n
&&
-8Tr(f_{\mu\nu}^{(i)}f_{\nu\sigma}^{(i)})
Tr(f_{\mu\rho}^{(j)}f_{\rho\
sigma}^{(j)})
+Tr(f_{\mu\nu}^{(i)}f_{\mu\nu}^{(i)})
Tr(f_{\rho\sigma}^{(j)}f_{\rho\sigma}^{(j)})\n
&&+Tr(f_{\mu\nu}^{(i)}\tilde{f}_{\mu\nu}^{(i)})
Tr(f_{\rho\sigma}^{(j)}\tilde{f}_{\rho\sigma}^{(j)}))
\label{blockin}
\eeqa
where $\tilde{f}_{\mu\nu}=\epsilon_{\mu\nu\rho\sigma}f_{\rho\sigma}/2$
and
\beq
\tilde{b}_8(f)=
{2\over 3}(Tr(f_{\mu\nu}f_{\nu\lambda}f_{\lambda\rho}f_{\rho\mu})
+2Tr(f_{\mu\nu}f_{\lambda\rho}f_{\mu\rho}f_{\lambda\nu})
-{1\over 2}Tr(f_{\mu\nu}f_{\mu\nu}f_{\lambda\rho}f_{\lambda\rho})
-{1\over 4}Tr(f_{\mu\nu}f_{\lambda\rho}f_{\mu\nu}f_{\lambda\rho})).
\eeq
In eq.(\ref{blockin}), we have kept axion type interactions also.
Note that the $\tilde{b}_8(f)=0$ for an (anti)instanton configuration.
So the potential between an instanton and an (anti)instanton is
\beq
{3\over 2 r^8}
(-8Tr(f_{\mu\nu}^{(i)}f_{\nu\sigma}^{(i)})
Tr(f_{\mu\rho}^{(j)}f_{\rho\sigma}^{(j)})
+Tr(f_{\mu\nu}^{(i)}f_{\mu\nu}^{(i)})
Tr(f_{\rho\sigma}^{(j)}f_{\rho\sigma}^{(j)})
+Tr(f_{\mu\nu}^{(i)}\tilde{f
}_{\mu\nu}^{(i)})
Tr(f_{\rho\sigma}^{(j)}\tilde{f}_{\rho\sigma}^{(j)})
).
\label{instint}
\eeq

Here we can apply the low energy approximation such as
\beqa
&&Tr(f_{\mu\nu}^{(i)}f_{\mu\nu}^{(i)})\rightarrow
{C^4\over l_{NC}^4}\int d^4xtr([D^i_{\mu},D^i_{\nu}][D^i_{\mu},D^i_{\nu}])
={l_{NC}^4\over \pi^2}, \n
&&Tr(f_{\mu\nu}^{(i)}\tilde{f}_{\mu\nu}^{(i)})\rightarrow
{C^4\over 2l_{NC}^4}\int d^4xtr\epsilon_{\mu\nu\rho\sigma}
([D^i_{\mu},D^i_{\nu}][D^i_{\rho},D^i_{\sigma}])
=\pm{l_{NC}^4\over \pi^2}, \n
&&Tr(f_{\mu\nu}^{(i)}f_{\nu\rho}^{(i)})\rightarrow
{C^4\over l_{NC}^4}\int d^4xtr([D^i_{\mu},D^i_{\nu}]
[D^i_{\nu},D^i_{\rho}])={l_{NC}^4\over 4\pi^2}\delta_{\mu\rho}
\label{approx}
\eeqa
where $D^i_{\mu}$ denotes the covariant derivative
of the instanton background which is localized at the $i$-th block.
So the interactions eq.(\ref{instint})
can be interpreted due to the exchange of
dilaton, axion and graviton.
We have found that the potential between two instantons vanish due to
their BPS nature.
On the other hand,
the following potential is found between an instanton
and an anti-instanton
\beq
-{3\over \pi ^4}{l_{NC}^8\over r^8}.
\label{gravin}
\eeq

There is no reason to believe the above approximation
when $b<l_{NC}$.
In this case the interactions between an instanton and
an anti-instanton is well described by the gauge fields which are low energy
modes of IIB matrix model. They are close to diagonal degrees of freedom
in IIB matrix model.
Their contribution can be estimated by gauge theory.
The one loop effective potential can be calculated by gauge theory
certainly when
$b<<l_{NC}$
\beq
\Gamma = -{C^4\over 2(4\pi )^2b^4}\int d^4x b_8
\eeq
where we have assumed $b\rho >> C$.
The above
 expression is estimated
as follows:
\beqa
\Gamma &=&-{144C^4\over \pi^2b^4}\int d^4x {\rho^4\over ((x-y)^2+\rho^2)^4}
{\rho^4\over ((x-z)^2+\rho^2)^4}\n
&&\sim -{24\rho^4C^4\over  r^{8} b^{4}}
\label{gaugein}
\eeqa
where $r=|y-z|$ is assumed to be much larger than $\rho$.
We note that eq.(\ref{gaugein}) falls off with the identical power for
large $r$
with eq.(\ref{gravin}).
On the other hand when $b>>l_{NC}$, the standard gauge theory description
is no longer valid since we have to take account of the noncommutativity.
In fact we have argued that the block-block interaction gives us the correct
result.
The purpose of this section is to underscore these arguments
and explain that the contributions of nonplanar diagrams in noncommutative
gauge theory indeed reproduce the block-block interactions.
We also provide a more accurate estimate of the crossover scale.

For this purpose, we
evaluate the leading term of the effective action by using the
plane-wave basis
$exp(ik\cdot \hat{x})$ just like the preceding section.
\beqa
W    &=&-{1\over n} \sum_k
Tr\left(exp(-ik\cdot \hat{x})\frac{1}{P^2}F_{\mu\nu} \frac{1}{P^2}
_{\nu\lambda}
             \frac{1}{P^2}F_{\lambda\rho}\frac{1}{P^2}F_{\rho\mu}
exp(ik\cdot \hat{x})\right) \n
       &~& -{2\over n} \sum_k Tr \left(exp(-ik\cdot \hat{x})\frac{1}{P^2}
F_{\mu\nu} \frac{1}{P^2}F_{\lambda\rho}
             \frac{1}{P^2}F_{\mu\rho}\frac{1}{P^2}F_{\lambda\nu}
exp(ik\cdot \hat{x})\right) \n
       &~&+\frac{1}{2n} \sum_k
Tr\left(exp(-ik\cdot \hat{x})\frac{1}{P^2}F_{\mu\nu}
\frac{1}{P^2}F_{\mu\nu}
             \frac{1}{P^2}F_{\lambda\rho}\frac{1}{P^2}F_{\lambda\rho}
exp(ik\cdot \hat{x})\right) \n
      &~&+\frac{1}{4n}\sum_k
Tr\left(exp(-ik\cdot \hat{x})\frac{1}{P^2}F_{\mu\nu}
\frac{1}{P^2}F_{\lambda\rho}
             \frac{1}{P^2}F_{\mu\nu}\frac{1}{P^2}F_{\lambda\rho}
exp(ik\cdot \hat{x})\right).
\eeqa
The above expression is calculated  as follows:
\beqa
W&=&n({1\over 2\pi B})^2\int d^4k
({1\over k^2+(Bb)^2})^4exp(iC^{\mu\nu}k_{\mu}l_{\nu})\n
&&\sum_{p,p',q}{3\over 2 }(
-8tr(f_{\mu\nu}^{(i)}(p)f_{\nu\sigma}^{(i)}(p'))
tr(f_{\mu\rho}^{(j)}(q)f_{\rho\sigma}^{(j)}(q'))
+tr(f_{\mu\nu}^{(i)}(p)f_{\mu\nu}^{(i)}(p'))
tr(f_{\rho\sigma}^{(j)}(q)f_{\rho\sigma}^{(j)}(q'))\n
&&+tr(f_{\mu\nu}^{(i)}(p)\tilde{f}_{\mu\nu}^{(i)}(p'))
tr(f_{\rho\sigma}^{(j)}(q)\tilde{f}_{\rho\sigma}^{(j)}(q')))
\label{blockgau}
\eeqa
where $l=p+p'=q+q'$.
Here we have assumed that the external momenta are small compared to
the noncommutativity scale. Hence we have dropped the phase which
only depends on the external momenta.
Note that eq.(\ref{blockgau}) contains the phase factor
$exp(iC^{\mu\nu}k_{\mu}l_{\nu})$ due to noncommutativity.

This phase becomes non-trivial when $|k|\sim 1/|l|l_{NC}^2$.
We can choose $1/|l|\sim \rho$ in this case.
Note that we have another scale $|k|\sim b/l_{NC}^2$ which is
associated with the propagator.
When $b > \rho >>l_{NC}$,
we show that eq.(\ref{blockgau}) can be
simply understood in terms of the block-block interactions.
It is evaluated as follows:
\beqa
W&=&n({1\over 2\pi B})^2{1\over (Bb)^8}\int d^4k
exp(iC^{\mu\nu}k_{\mu}l_{\nu})\n
&&\sum_{p,p',q}{3\over 2}(
-8tr(f_{\mu\nu}^{(i)}(p)f_{\nu\sigma}^{(i)}(p'))
tr(f_{\mu\rho}^{(j)}(q)f_{\rho\sigma}^{(j)}(q'))
+tr(f_{\mu\nu}^{(i)}(p)f_{\mu\nu}^{(i)}(p'))
tr(f_{\rho\sigma}^{(j)}(q)f_{\rho\sigma}^{(j)}(q'))\n
&&+tr(f_{\mu\nu}^{(i)}(p)\tilde{f}_{\mu\nu}^{(i)}(p'))
tr(f_{\rho\sigma}^{(j)}(q)\tilde{f}_{\rho\sigma}^{(j)}(q')))\n
&=&n^2{1\over (Bb)^8}\n
&&\times \sum_{p,q}{3\over 2}(
-8tr(f_{\mu\nu}^{(i)}(p)f_{\nu\sigma}^{(i)}(-p))
tr(f_{\mu\rho}^{(j)}(q)f_{\rho\sigma}^{(j)}(-q))
+tr(f_{\mu\nu}^{(i)}(p)f_{\mu\nu}^{(i)}(-p))
tr(f_{\rho\sigma}^{(j)}(q)f_{\rho\sigma}^{(j)}(-q))\n
&&+tr(f_{\mu\nu}^{(i)}(p)\tilde{f}_{\mu\nu}^{(i)}(-p))
tr(f_{\rho\sigma}^{(j)}(q)\tilde{f}_{\rho\sigma}^{(j)}(-q)))\n
&=&{1\over (Bb)^8}\n
&&\times {3\over 2}(
-8({1\over 2\pi C})^2\int d^4xtr(f_{\mu\nu}^{(i)}(x)f_{\nu\sigma}^{(i)}(x))
({1\over 2\pi C})^2\int d^4ytr(f_{\mu\rho}^{(j)}(y)f_{\rho\sigma}^{(j)}(y))\n
&&+({1\over 2\pi C})^2\int d^4xtr(f_{\mu\nu}^{(i)}(x)f_{\mu\nu}^{(i)}(x))
({1\over 2\pi C})^2\int d^4y
tr(f_{\rho\sigma}^{(j)}(y)f_{\rho\sigma}^{(j)}(y))\n
&&+({1\over 2\pi C})^2\int
d^4xtr(f_{\mu\nu}^{(i)}(x)\tilde{f}_{\mu\nu}^{(i)}(x))
({1\over 2\pi C})^2\int
d^4ytr(f_{\rho\sigma}^{(j)}(y)\tilde{f}_{\rho\sigma}^{(j)}(y))).
\eeqa
Using the large instanton size approximation eq.(\ref{approx}),
we indeed reproduce the result of eq.(\ref{gravin}).

On the other hand, we can neglect the phase factor
$exp(iC^{\mu\nu}k_{\mu}l_{\nu})$ when $b < \rho$.
With this approximation, we obtain:
\beqa
W&=&n({1\over 2\pi B})^2\int d^4k
({1\over k^2+(Bb)^2})^4\n
&&\sum_{p,p',q}{3\over 2 }(
-8tr(f_{\mu\nu}^{(i)}(p)f_{\nu\sigma}^{(i)}(p'))
tr(f_{\mu\rho}^{(j)}(q)f_{\rho\sigma}^{(j)}(q'))
+tr(f_{\mu\nu}^{(i)}(p)f_{\mu\nu}^{(i)}(p'))
tr(f_{\rho\sigma}^{(j)}(q)f_{\rho\sigma}^{(j)}(q'))\n
&&+tr(f_{\mu\nu}^{(i)}(p)\tilde{f}_{\mu\nu}^{(i)}(p'))
tr(f_{\rho\sigma}^{(j)}(q)\tilde{f}_{\rho\sigma}^{(j)}(q')))\n
&=&-{C^4\over 2(4\pi )^2b^4}\int d^4x b_8.
\eeqa
In this way, we reproduce the ordinary gauge
theory result eq.(\ref{gaugein}).

\section{Conclusions}
In this paper we have proposed a bi-local field representation
of noncommutative field theories.
In the momentum eigenstate representation, the momenta of fields
can become much larger than the noncommutative scale $l_{NC}$.
However
we can no longer regard these large momentum degrees
of freedom as ordinary local fields
since the wave functions become noncommutative.
Due to the noncommutativity of space-time,
the wave functions with large momenta are more naturally interpreted as
translation operators  and hence those degrees of freedom
are interpreted as bi-local fields.
In the $n \times n$
matrix reguralization of noncommutative field theories,
there are $n^2$ degrees of freedom and this number is much larger
than $n$ which is the degrees of freedom of ordinary local field with
UV cutoff $l_{NC}^{-1}$.
This implies that noncommutative field theories
contain infinitely many particles.
It is interesting to investigate the possibility that
these infinite degrees of freedom may pile up to strings.
\footnote{see also \cite{Mkato}}
\par
We have also calculated long range interactions in noncommutative
field theories. In the conventional approach, they manifest as
IR singular behaviors of nonplanar diagrams.  We have seen that
this feature can be
naturally understood in terms of
block-block interactions in the matrix model picture.
This type of interactions are characteristic to matrix models,
not restricted to twisted reduced models which are directly
related to noncommutative field theories.
In the case of type IIB matrix model,
we have shown \cite{IKKT} the appearance of block-block
interactions in generic backgrounds and interpreted them as propagations of
massless supergravity multiplets.
These interactions differ from those of the ordinary gauge theory
such as $D=4$ ${\cal N}=4$ super Yang-Mills field theory
as is explained in the preceding section.
This point demonstrate the richness of matrix models over the corresponding
field theories and it is one of advantages to consider
IIB matrix model as a constructive formulation of superstring.

\begin{center} \begin{large}
Acknowledgments
\end{large} \end{center}
This work is supported in part by the Grant-in-Aid for Scientific
Research from the Ministry of Education, Science and Culture of Japan.
We would like to thank M. Hayakawa and T. Tada for discussions.

\setcounter{equation}{0}

\end{document}